\documentstyle[12pt,aaspp4,epsfig]{article}

\def\deg{\ifmmode^\circ\else$^\circ$\fi}

\def\mic{~$\mu$m}

\def\eg{{\it e.g.}}

\def\et{et al.}
\def\myr{\mbox{\Msun yr$^{-1}$}}
\def\arcs{\ifmmode {''}\else $''$\fi}
\def\arcm{\ifmmode {'}\else $'$\fi}
\def\parcs{\sa=.07em \sb=.03em
     \ifmmode $\rlap{.}$^{\scriptscriptstyle\prime\kern -\sb\prime}$\kern -=
\sa$
     \else \rlap{.}$^{\scriptscriptstyle\prime\kern -\sb\prime}$\kern -\sa\=
fi}
\def\parcm{\sa=.08em \sb=.03em
     \ifmmode $\rlap{.}\kern\sa$^{\scriptscriptstyle\prime}$\kern-\sb$
     \else \rlap{.}\kern\sa$^{\scriptscriptstyle\prime}$\kern-\sb\fi}
\def\mpc3{\mbox{Mpc$^{-3}$}}
\def\Msun{M$_{\odot}$}
\def\Myr{\Msun/yr}
\def\kp{{\rm K}$^{\prime}$}
\def\lya{{\rm Ly}$\alpha$}

\def\han {\mbox{{\rm H}$\alpha$}}

\def\ha{\han}

\def\pb {{\rm P}$\beta$}

\newcommand{\mgii}{\mbox{Mg {\sc ii}}}
\newcommand{\nii}{\mbox{[N {\sc ii}]}}

\newcommand{\civ}{\mbox{C {\sc iv}}}

\newcommand{\oii}{\mbox{O {\sc ii}}}

\def\spose#1{\hbox to 0pt{#1\hss}}
\def\simlt{\mathrel{\spose{\lower 3pt\hbox{$\mathchar"218$}}
     \raise 2.0pt\hbox{$\mathchar"13C$}}}
\def\simgt{\mathrel{\spose{\lower 3pt\hbox{$\mathchar"218$}}
     \raise 2.0pt\hbox{$\mathchar"13E$}}}
\def\lsim{\rlap{$<$}{\lower 1.0ex\hbox{$\sim$}}}
\def\gsim{\rlap{$>$}{\lower 1.0ex\hbox{$\sim$}}}

\begin{document}

\title{Near-Infrared Observations of the Environments of Radio Quiet QSOs
  at z$\simgt$1}

\author { Harry I. Teplitz \altaffilmark{1} \altaffilmark{2}}
\affil{
Goddard Space Flight Center, Code 681, GSFC, Greenbelt, MD 20771 \\
hit@binary.gsfc.nasa.gov}

\author {Ian S. McLean, Matthew A. Malkan}
\affil{Department of Physics \& Astronomy, Division of Astronomy \\
University of California at Los Angeles \\
Los Angeles, CA  90095-1562 \\
 malkan@astro.ucla.edu, mclean@astro.ucla.edu}
\altaffiltext{1}{NOAO Research Associate}
\altaffiltext{2}{Previous affiliation:  Department of Physics \&
Astronomy, UCLA}

\begin{abstract}

  We present the results of an infrared survey of QSO fields at z=0.95,
  0.995 and 1.5.  Each $z< 1$~field was imaged to typical continuum
  limits of J=20.5, \kp=19 ($5\sigma$), and line fluxes of $1.3\times
  10^{-16}$~ergs/cm$^2$/s ($1\sigma$) in a 1\% interference filter.
  16 fields were chosen with $z\sim 0.95$~targets, 14 with $z \sim
  0.995$~and 6 with $z\sim 1.5$.  A total area of 0.05 square degrees
  was surveyed, and two emission-line objects were found.  We present
  the infrared and optical photometry of these objects.  Optical
  spectroscopy has confirmed the redshift of one object (at z=0.989)
  and is consistent with the other object having a similar redshift.
  We discuss the density of such objects across a range of redshifts
  from this survey and others in the literature.  We also present
  number-magnitude counts for galaxies in the fields of radio quiet
  QSOs, supporting the interpretation that they exist in lower density
  environments than their radio loud counterparts.  The J-band number counts
  are among the first to be published in the J=16--20
  range.

\end{abstract}

\keywords{galaxies : individual --- galaxies : evolution --- cosmology :
observation --- infrared : galaxies}

\section {Introduction}

Current estimates of the global star formation rate show a steep rise
in activity out to a redshift of 1 (Lilly et al. 1996).  Half of the
stars in the current universe may have formed since z=1.  High
redshift star-forming galaxies are also seen to be highly biased
tracers of the mass distribution of the universe (Steidel et al.
1998).  As a result of these two observations, significant effort has
been devoted in recent years to the search for galaxy clusters at
redshifts beyond z=1 (Dickinson 1997, Donahue et al. 1998, etc.)
Narrow-band searches for the \ha~emission line have successfully
identified star forming galaxies at $z>2$~(Malkan, Teplitz \& McLean
1995, 1996; Teplitz, Malkan, \& McLean 1998, hereafter TMM98; and
Mannucci, Thompson, Beckwith, and Williger 1998, hereafter MTBW98),
and searches for [OII] and [OIII] have been used at $z\sim
0.9$~(MTBM98) and $z > 3$~(Teplitz, Malkan, and McLean 1999; hereafter
TMM99).  In this paper we apply this imaging infrared technique to the
search for $z\sim 1$~galaxies.

It takes patience to find the relatively rare galaxies at z=1 simply by
covering large areas of random field.  Instead, we have chosen 
the popular tactic of looking for galaxies where they are most likely
to be:  near other galaxies or QSOs.    In particular, we search
fields containing radio-loud or radio-quiet quasars, and fields containing
absorption line systems in QSO lines of sight.  All of these objects
have previously been associated with the presence of groups of galaxies.

Yee \& Green (1987) and later Ellingson, Yee, \& Green (1991) have
examined galaxy clusters as the environment of quasars, though their
searches extend only out to z=0.7.  They find significantly more
bright QSOs in clusters at higher redshifts, indicating that quasars
may be good markers of galaxy clusters.  Similarly, large numbers of
excess galaxies were found in Radio Loud QSO fields at $1<z<2$~by Hall
\& Green (1998; see also Hall, Green \& Cohen 1998).  At $z\sim 2$,
Arag\'{o}n-Salamanca \et~(1996) find an excess of K-band selected
galaxies in the environments of QSOs.  Also at $z>2$, Hutchings (1995)
finds excess R-band selected galaxies in the fields of both radio-loud
and radio-quiet QSOs.  In similar searches, high redshift radio
galaxies are also seen as markers of other clustered galaxies
(Dickinson 1997).

Absorption-line systems have also been studied in detail for their
galaxy content. Galaxies have been identified as responsible for
\mgii\ and \civ\ absorption out to $z\sim 1$~(\eg~Steidel \et, 1997).
Arag\'{o}n-Salamanca \et~(1994) have demonstrated excess K-selected
galaxies in fields containing \civ~absorbers.  Possible Damped
\lya~Absorption (DLA) systems have been identified past $z=3$
(Djorgovski 1996, Moller \& Warren 1998).  The presence of galaxies as
the absorbers make such systems good markers for other galaxies.
However, the small impact parameters will often make the absorbers
themselves hard to see except with very high resolution.

We have chosen to search for \ha~as an indicator of star formation in
normal galaxies.  It is strong in nearby spirals, with
equivalent widths of 50\AA\ or more in the actively star-forming
galaxies (Kennicutt, 1983).  There are several methods of identifying
on-going star formation, and determining the Star Formation Rate (SFR)
with various degrees of accuracy.  Emission may either be observed
from the UV continuum of young stars, from the ionized gas in HII
regions, or re-radiated in the FIR by dust.  \ha~emission, at
6563\AA, can be observed from nebular emission in the star-forming HII
regions.  This emission line has the advantage of relatively long
wavelength, making it less susceptible to extinction than the UV
continuum.  There is a concern that the emission actually observed is
a combination of lines, but the \nii~6584\AA\ line is found to be weak in most
cases, at most 10--30\% of \ha~(Kennicutt \& Kent, 1983).

Kennicutt (1983) calculates the SFR based on \ha~luminosity to be
\begin{equation}
\label{ref:  Kennicutt SFR}
\mbox{SFR(total)} = \frac{L(\han)}{1.12\times 10^{41}\mbox{ergs~s}^{-1}}
        \mbox{\Msun} \mbox{yr}^{-1}
\end{equation}
for an Initial Mass Function (IMF) that is effectively that of
Salpeter, with $\psi(m) \propto m^{-2.5}$, and an upper mass
cutoff of 100\Msun.  Similar estimates can be made from observations
of the \oii~emission line (Kennicutt 1992, but see Hammer \et, 1997).

The \ha~emission line has also been used to measure the SFR in high redshift
galaxies selected by other methods.  For example, Glazebrook et al.~(1998)
have spectroscopically measured \ha~in 13 $z\sim 1$~galaxies selected
from the Canda-France Redshift Survey (see Lilly et al. 1996).  They find 
that $z\sim 1$~galaxies may have SFRs three times higher than would be
inferred from rest-frame UV observations.  

\section{Observations}

All IR observations in the $z \sim 1$ survey were taken on the Shane
3m telescope at Lick Observatory on Mount Hamilton, with the UCLA Two
Channel Near-IR Camera, colloquially known as ``Gemini" for its twin
detectors, (McLean, 1994).  Use of this instrument for deep IR imaging
has been discussed in McLean \& Teplitz (1997; hereafter MT97) and
Larson \et~(1997).  The observations of the six $z\sim 1.5$ fields
were taken at the 10m Keck I telescope using the Near IR Camera, NIRC
(Matthews \et, 1994).  For a discussion of filter transmission in the
NIRC instrument see TMM98.

The UCLA 2-Channel Near-IR Camera (see also McLean \et, 1993, 1994)
has two independent infrared arrays, one $256\times 256$ pixel NICMOS
3 HgCdTe device and one $256\times 256$ pixel SBRC InSb detector.  The
camera is a low-noise, high-throughput system.  To optimize
performance, the instrument employs a variety of readout modes,
including Correlated-Double-Sampling, Fowler Sampling (see Fowler \&
Gatley 1990), a multiple-reads-per-pixel mode, and subarray readouts.
Typical deep integrations, like those we employ in this project,
utilize Fowler sampling, with 8--32 multiple reads. The plate scale
for the detectors is 0.70\arcs and $\sim 0.68$\arcs / pixel for the
short- and long-wavelength channels.  These scales correspond to a
field of view of $\sim$3\arcm $\times$ 3\arcm.  We chose a dichroic
beam splitter which directs light longward of 2\mic~to the SBRC array.
Our typical exposure times were three coadds of 50 seconds each in the
J band, 18 coadds of 8 seconds in \kp, and a single 240 second
exposure in the narrow band.

Each channel has a filter wheel containing 25.4 mm diameter
transmission and blocking filters.  In addition to the standard J,H,
and \kp~filters, this project utilized two narrow-band interference
filters.  The filters were purchased from Barr Associates in Westford,
Massachusetts.  One was ``off the shelf", while the other was custom
made for this project.  Table \ref{tab: 4.2} summarizes the
properties of each filter, as described below.  The first narrow-band
filter is the standard filter at 1.28\mic, usually used to measure the
Paschen $\beta$~(\pb)~transition.  It is 141\AA\ wide, which
corresponds to $\Delta \lambda / \lambda = 1.1\%$.  
Standard star observations show that the ratio of \pb~filter
transmission to standard J filter transmission is $1:22\pm 1$.  The
wavelength of the \pb~filter corresponds to the redshifted wavelength
of \ha~at at $z=0.95$.

%It was purchased
%with a guarantee of 60\% transmission ($T_n$).  
%Assuming a flat spectrum across the
%relatively small waveband of the filters, the relative measured flux
%density transmitted by the narrow filter, $f_{n}$, will be the product
%of its width and transmission:
%\begin{equation}
%\frac{f_{n}}{f_{b}} = \frac{\Delta \lambda_{n} T_{n}}{\Delta \lambda_{b} 
%T_{b}}
%\end{equation}
%The filter is 1/20.3 the width of the J filter, so if the particular J
%filter installed in the UCLA Camera has 70\% peak transmission as
%promised by the manufacturer, the \pb~ filter meets its specification,
%and exceeds its specification with $T_{n}=65\%$.  

The second narrow band filter was custom-designed for this project.
The filter is centered on 1.3095\mic, to probe a redshift range higher
than that of the \pb~filter.  The choice was constrained by the
atmospheric absorption at the red end of the J passband.  At
$z=0.995$, \ha~redshifts into this filter.  The filter was produced
using a new oxide manufacturing process, in order to maximize
throughput and cryogenic performance.  The filter transmission is
fairly flat around the peak of 78\%, centered on 1.3095\mic, with a
width of 0.0130\mic, as verified in a cryogenic test.  The
transmission was confirmed by standard star observations which show
that the ratio of transmitted flux compared to the J filter is
$1:19\pm 1$.  
%The ratio of filter widths is 1:21.9, so given the J
%filter transmission confirmed by the \pb~filter, the manufacturer's
%transmission measurement is verified.

Targets were selected from a search of NASA/IPAC Extragalactic
Database, in fields containing QSOs (either radio loud and radio
quiet) or absorption line systems (either DLA or or metal line
absorbers).  Objects below -30\deg\ were not considered as Keck
targets, nor south of -5\deg\ as Lick targets.  Subject to these
constraints, we then selected fields at redshifts within $\pm 0.3
\%$~of the central redshift of the filters.

The data were reduced following the procedures outlined in TMM98 and
MT97 which will only be summarized here.  We obtained images in a
sequence of ``dithered'' exposures, offsetting the telescope between
exposures in a 3$\times$3 grid typically spaced by 10--20 pixels.  The
individual frames were divided by a twilight flat or dome flat (as
available). Then a running median sky frame (created from the nine
exposures taken closest in time to each image) was subtracted.  When
no such flat was available, the running median was divided into each
image and sky subtraction was left to aperture photometry.  Objects
were identified using the SExtractor (Bertin \& Arnouts 1996)
software.  Photometry was performed using apertures of 2.5 times the
seeing disk.  The same aperture was applied to broad and narrow-band
exposures. Similarly, the same photometric aperture was applied to
both J and \kp~frames.  Photometric errors were estimated from aperture
photometry performed on random positions in the frame.  Errors in the
narrow-minus-broad band color were estimated from a Monte Carlo
simulation of the ratio of numbers with large gaussian errors in order
to define the confidence intervals.  (See TMM98 for details).

Follow-up optical imaging and spectroscopy was obtained with the Low
Resolution Imaging Spectrometer (LRIS), which has a 2048$\times$2048
pixel CCD detector, with a field of view of 5\arcm~by 7\arcm~(Oke \et,
1995).  Spectra were taken with the 300 lines/mm grating, giving a
dispersion of 2.48\AA~pixel$^{-1}$, and spectral resolution (FWHM) of
4.5 pixels.  These data were reduced with standard IRAF procedures.

\section{Results}

There were 43 known targets meeting our criteria.  Of these possible
fields, we imaged 30 during the 11 clear nights available for this
project (see Table \ref{tab: z1obs}).  Table \ref{tab: z1 qso phot}
lists QSO photometry for each field, and estimated optical magnitudes
from the NED database.  In each field, we obtained simultaneous J and
\kp~imaging at integration times designed to detect $\sim$L* galaxies
at the appropriate redshifts.  Fields observed at high signal-to-noise
or that seemed otherwise promising were planned for narrow-band
followup.  Table \ref{tab: z1nbobs} lists narrow band observations and
inferred limiting SFR.  Six fields observed with z=1.5 ``signposts''
are also listed in Table \ref{tab: z1pt5nbobs}.

\subsection{Discussion of Individual Fields}

In each field, the errors in the Broad--Narrow colors were calculated
as discussed, above.  Objects with narrow-band excesses lying at above
the 99\% confidence interval are considered possible detections.  Two
apparent \han-emitters were detected in this survey.  They are listed
in Table \ref{tab: z1 nb detections}, along with their inferred SFR
and broad-band photometry.  

Our detection limits were typically J=20.5 (5$\sigma$) and SFR = 12 \myr
($2\sigma$).  This depth should be sufficient to pick up vigorously
star forming galaxies at z=1.  The GISSEL96 (see Bruzual \& Charlot 1993) 
models predict that an
evolved L* galaxy at z=1 will have J=20.5.  Adding a strong ongoing
episode of star formation encompassing 10\% of the galaxy's mass over
1Gyr would lead to a SFR=10\myr.

The survey has covered approximately 190 square arcminutes (in the
highest SNR regions).  This is sufficient to constrain the size of a
moderately high-redshift galaxy population with SFR's at the
sensitivities we have achieved.

\subsection*{0107-025}
This field was of particular interest because there are two known QSOs
within one UCLA Camera field of view with emission redshifts that
place \ha~in the \pb~filter (see Figure \ref{fig: im0107}). However no
new line-emitting objects were detected.  The two QSOs are clearly
detected in the narrow band.  Figure \ref{fig: comphist0107} compares
the number of galaxies in this field to the average J number counts
for all the broad band fields.  Since little J-band field data are
availabe at these magnitudes (see Section 5), we use our averaged
number counts.   There is a possible excess in several
bins at $J>19$, about 1$\sigma$ above the average per bin.

\subsection*{2145+067}

Figure \ref{fig: im2145} shows the broad-band image of this field.  
No statistical excess of galaxies 
has been detected (see Figure \ref{fig: comphist2145}).  A drawback to
its consideration as a target is the bright star in the southern
region of the field.  As a result, we decided against it in the first
set of narrow-band observations in August of 1995.  However, we did
select it as a target the following year, and observed it in October,
1996.

One object in this field shows an apparent $\sim 3\sigma$ excess in
the narrow band filter (see Figure \ref{fig: cm2145}).  Its
$\Delta$m$=1.2$~yields an inferred SFR=53\Myr, assuming the line is
\ha~at $z=0.995$.  The inferred equivalent width would be 132\AA~or
2\% $EW/\lambda$.  The object lies 45\arcs~from the QSO.  It has a FWHM
of 4.2\arcs, compared to a seeing disk of 2.2\arcs.  Our photometry
shows the galaxy to have $I=22.5$, $J=19.2$, and \kp$=18.6$.  Optical
CCD imaging of this field as a followup of HST QSO fields (Kirhakos
\et, 1994) did not detect this object, down to a limit of $g\sim=22$,
consistent with our I band detection.  These colors are generally
consistent with a $z\sim 1$~galaxy, where the 4000/3646\AA~break lies
in the I band.  Assuming passive evolution, this galaxy would be
somewhat brighter than $L_*$~today.

The line-emitting galaxy in this field is located 14\arcs~west of a
bright star.  This initially suggested that the detection could
actually be a ghost.  To test this, an image was taken through the
same filter of a bright star in a relatively uncrowded field.  Two
ghosts were identified to the SE of the star, as a distinctive
``double".  No ghost image was detected at the position of the
possible \ha-emitter.  Also, the strong detection of the object in
both the J and \pb~filters makes the risk of a ghost less likely.

We subsequently detected this object with LRIS imaging and spectroscopy,
demonstrating that it is real and not a ghost.  We do not find any
emission lines to confirm the object's redshift.  In a future paper,
we will present an analysis of cross-correlation of this spectrum with
galaxy spectra at various redshifts (see also Cohen et al. 1998).

\subsection*{LBQS 2350-0132}
A single, strong \han-emitting object is detected in this field (see
Figures \ref{fig: im2350} and \ref{fig: cm2350}).  It is shows
\ha~emission with an EW of 1.3\%, or SFR=26.0\myr. The line flux is
$5.1\times 10^{-16}$ergs/cm$^2$/sec., and the equivalent width is
86\AA\ in the restframe. The object is clearly extended with
FWHM=3.0\arcs~compared to a seeing disk of 1.5\arcs.  It lies
47\arcs~from the QSO (400kpc at z=0.993).

The J-\kp=1.3 color of this object is blue for an evolved elliptical
at this redshift, but could be consistent with a starburst.  Its
\kp~continuum magnitude is roughly consistent with an L* galaxy at
this redshift, suggesting that this is a secondary starburst, not the
initial formation of the galaxy.  Photometry with LRIS shows the
galaxy to have $I=21.9$, still fairly blue for galaxies at that
redshift.  This same field was observed by Kirhakos \et (1994).  They
do not detect this object down to a limit of $g\sim 22$, again
consistent with our photometry.

The number counts for this field do not show any excess of faint
galaxies (see Figure \ref{fig: comphist2350}).  Similarly, the reddest
objects in the frame are not clustered near the line-emitter, nor near
each other.  A single very red object (J-\kp=2.8) lies $\sim
45$\arcs~north of the detection, though it falls in a lower SNR region
of the image.

We have obtained LRIS imaging and spectroscopy of this field.  The
redshift inferred for the narrow-band detection is confirmed to be
$z=0.989 \pm 0.001$~by the presence of the [OII] emission line in the
optical spectrum (see figure \ref{fig: spec2350}).  The redshift
difference between the QSO and the galaxy (along the line of sight)
corresponds to 4.2 Mpc at z=0.99, which is large to suggest no
physical connection with the signpost.  The [OII] line is unresolved
at the resolution of the spectrum, which constrains the intrinsic
width to be no greater than 300 km/s.  The line flux is $1.9\times
10^{-16}$ ergs/cm$^2$/s.  The equivalent width of the [OII] line is
80\AA~in the restframe, or 93\% of the equivalent with of the \ha+NII
complex.  Kennicutt (1992) finds that most low redshift star forming
galaxies have EW([OII]):EW(\ha) ratios of 0.40, though there is
considerable scatter for objects with EW(\ha)$>40$\AA, such as this
one.  Kennicutt also comments that Seyfert 2 galaxies are found to
have high EW([OII]):EW(\ha) ratios while Seyfert 1's have low ratios.
Thus it is possible that this candidate object contains a Seyfert 2
nucleus, though the line ratio does not preclude a purely stellar
origin. The angular extent of the continuum image argues against a
broad-line object (Seyfert 1). Even a narrow-line AGN (Seyfert 2)
would be expected to produce some detectable Mg II emission, which
should have fallen in a good region of our spectrum, but is not seen.
The $3\sigma$~limit at the wavelength of redshifted Mg II is $6\times
10^{-18}$~ergs/cm$^2$/sec.

\section{Emission Line Galaxies at $z\sim 1$}

We have detected one definite and one probable \han-emitting galaxy at
$z\ge0.95$.  This corresponds to a surface density of 0.01 galaxies
per square arcminute.  For comparison at z=3 in large ($\Delta z>0.4$)
redshift windows, field galaxies are detected, with SFR$\sim$8\myr, at
0.4--0.8 per square arcmin (Steidel \et, 1996).  

As a volume density, our two detections yield $4.7^{+6}_{-3}\times
10^{-4}$ galaxies/Mpc$^3$~(comoving).  This density is similar to that
for field \ha-emitting galaxies at this redshift, as determined by the
NICMOS parallel grism survey (McCarthy et al.  1999).  Table \ref{tab:
  halpha_surveys_tab} lists results of various \ha~surveys for
star-forming galaxies for different redshifts.  Comparing the results,
we find broad agreement in the densities of \ha-emitters, with a
higher density in absorption-line fields, as already noted in TMM98
and MTBW98.  We also find good agreement in the average star formation
rates at different redshifts.  This constant detected average SFR may
be the result of the bias imposed by a flux-limited survey (for
example TMM98 found fainter, less vigorously star forming objects by
going deeper).  However, the consistency in the average SFR of
galaxies brighter than the flux limit may be indicative of the
constant global star formation rate observed at $z>1$.  Lyman Break
galaxy searches (Steidel et al. 1998) and observations in the sub-mm
(c.f. Smail et al. 1998) have suggested that the global SFR does not
change much between redshifts of 1 and 5.

\section{Number-Magnitude Counts}
Our typical broad-band detection limits are J=20.5 and \kp=19.0,
corresponding to an evolved $L_*$~galaxy at $z\sim 1$.  In Figures
\ref{fig: jnc} and \ref{fig: knc} we plot the number-magnitude counts
in J and \kp.  For comparison, the number counts from the literature
are also plotted.  In both the J and \kp~bands, we find no statistical
excess of galaxies averaged over the total survey area in QSO fields.
It has been suspected for some time that radio-quiet QSOs (RQQs) at
most redshifts do not lie in high density environments, unlike
radio-loud QSOs (RLQs).  Most of the quasars in our sample are RQQs,
and so it is perhaps not surprising that our number counts agree with
the field.  Table \ref{tab:  jnc}~lists the number-magnitude counts
per square degree per magnitude for the J band.

Our J-band number counts are among the first wide-field number-counts
for any environment at deep J magnitudes.  The DENIS Survey (Mamon et
al., 1998) has provided J-band number counts down to J=15.  Deep
J-band counts will soon be available from the NICMOS parallel survey
(c.f. Teplitz et al. 1998, Yan et al. 1998), and counts at J fainter
than 20 have been measured by Bershady et al.  1998.  It is of
particular interest to have number counts in different IR wavebands,
in order to provide good field comparisons for the study of cluster
environments at different wavelengths (c.f. Stanford et al.  1997).

Our result that RQQs lie in low density environments is supported by
the conclusion reached by Croom \& Shanks (1998) for RQQs at $z<1.5$.
They find no significant clustering signal for galaxies down to $L_*$.
Their observations were in the $B_j$-band, sampling the restframe
near-UV, so a K-band survey provides significant new data.  Similarly,
both TMM98 and Mannucci et al. (1998) find few galaxies associated
with $z<3$~(mostly radio quiet) QSOs.

On the other hand, Jaeger et al. (1999) find a significant excess of
galaxies close to RQQs at $z\sim 1$~in R band observations.  We can
also consider the radial distribution of galaxies in reference to the
QSO for our fields in the J band (see Figure \ref{fig: radhist}).
This analysis is more sensitive to small scale clustering, that has no
impact on the average number counts over then entire area.  There is a
marginally significant excess of galaxies within $\sim 15$\arcs~of the
QSO ($\sim 120$~kpc at z=1).  We note however that there are a total
of only 66 galaxies in the three innermost bins, compared with a field
expectation of 47 galaxies, or an average excess of one galaxy per two
fields.  The field expectation the same, whether it is determined from
the average number counts over the entire survey area, or from
repeating the radial binning around random points in the fields.  Such
a small possible excess is only present at the $\sim 2\sigma$~level.
In addition, even accepting such an excess, it argues for RQQ
environments being poor groups of galaxies, not rich cluster.
However, the radial extent is generally consistent with the
results of TMM98, which found no roll-off in the radial distribution
of \ha-emitters within one NIRC field (38\arcs $\times$ 38\arcs).

The potentially excess galaxies detected in the broad-band, if they
are physically associated with the quasar signposts (within 2\% of
redshift), have a density which is higher than that of strong
\ha-emitters.  This effect would not be unexpected, if one assumes
that there is a large population of less actively star-forming
galaxies by a redshift of one.  The galaxies around these quasars
might be seen to be actively star-forming at a much earlier epoch.
For example, Hu et al. (1998) and Djorgovski et al. (1997) find high
densities of \lya-emitters around $z>4$~QSOs.  This could imply that
by $z\sim 1$, little active star formation is to be found in poor
groups.  By contrast, richer groups around QSOs at $z\sim 1.5$~are
seen to have a higher density of \ha-emitters (Hall 1998).  Indeed, by
$z\sim 0.4$, active star-formation is more often observed in rich
clusters (see Oemler 1992).

\section{Future Prospects}

This project has demonstrated that future narrow-band imaging surveys
can succeed in probing the $z \ge 1$ Universe, if sufficiently long
integration times are combined with new large-format infrared arrays.
At that point it will be reasonable to make blind searches of the
field, which will reveal if the targeted searches described here have
benefited much from galaxy clustering around quasars and their
absorbers.

\acknowledgements

We thank the Telescope Technicians at Lick Observatory and the
Observing Assistants and Instrument Specialists at the Keck
telescopes.  We also thank the members of the UCLA IR Detector Lab,
especially J. Canfield and N.  Magnone for enabling the many observing
runs with the UCLA IR Camera.  We thank E. Becklin, B. Zuckerman, and
P. Lowrance for aquiring a portion of the data on the 0500+019 field
We thank Matthew Bershady and Gary Mamon for making their number
counts available to us in electronic form; and Jonathon Gardner and G.
Williger for useful discussions.

This research has made use of the NASA/IPAC
Extragalactic Database (NED) which is operated by the Jet Propulsion
Laboratory, California Institute of Technology, under contract with
the National Aeronautics and Space Administration.  Some data
presented herein were obtained at the W.M. Keck Observatory, which is
operated as a scientific partnership among the California Institute of
Technology, the University of California and the National Aeronautics
and Space Administration.  The Observatory was made possible by the
generous financial support of the W.M. Keck Foundation.

\clearpage

\renewcommand{\arraystretch}{.5}
\begin{deluxetable}{llll}
\tablecolumns{4}

\tablecaption{Filters in the UCLA 2-Channel Camera}

\tablehead{
\colhead{Filter}&
\colhead{central $\lambda$ (\mic)}&
\colhead{$\Delta \lambda$ (\mic)}&
\colhead{Tranmission(\%)}
}

\startdata

J       & 1.2    & 0.285   & 70   \nl
\pb     & 1.28   & 0.01408 & 65    \nl
1.3095  & 1.3095 & 0.0130  & 78  \nl
H       & 1.65   & 0.3     & $\ge 70$\tablenotemark{a} \nl
\kp     & 2.1    & 0.35    & $\ge 70$\tablenotemark{a} 

\enddata
\tablenotetext{a}{manufacturer specification}
\label{tab: 4.2}
\end{deluxetable}
\renewcommand{\arraystretch}{1.0}

\clearpage
\renewcommand{\arraystretch}{.5}
\begin{deluxetable}{lcccc}

\tablecolumns{5}
 
\tablecaption{Broad Band Observations of $z\sim 1$ Targets}
\tablehead{
\colhead{Field}&
\colhead{date}&
\colhead{itime (sec)}&
\colhead{J 5$\sigma$}&
\colhead{\kp~5$\sigma$} 
}

\startdata

0107-025 & 08/04/95 & 2700 & 20.87 & 18.53 \nl 
0122-003 & 08/05/95 & 3300 & 20.76 & 18.60 \nl
0246+009 & 10/19/96 & 3600 & 20.41 & 19.20 \nl
0308+002 & 10/27/96 & 2700 & 20.40 & 18.49 \nl
0333+321 & 10/19/96 & 3600 & 19.92 & 19.23 \nl
0447-092 & 10/19/96 & 3600 & 20.52 & 18.96 \nl
0500+019 & 10/19/96 & 4500 & 20.05 & 18.98 \nl
4c +22.21 & 03/13/96 & 3600 & 20.67 & 19.38 \nl
0743-006 & 03/13/96 & 3700 & 19.83 & 18.63 \nl
CBS 0076  & 10/27/96 & 3600 & 21.11 & 19.25 \nl
0903+474 & 03/13/96 & 3600 & 20.41 & 19.05 \nl
1050+542 & 03/12/96 & 3600 & 20.74 & 18.96 \nl
1231+163 & 03/12/96 & 3600 & 20.79 & 19.29 \nl
1235+112 & 03/12/96 & 3600 & 20.75 & 19.08 \nl
1244+324 & 05/20/95 & 3600 & 20.81 & 19.24 \nl
1306+296 & 05/06/96 & 3600 & 20.47 & 18.97 \nl
1307+296\tablenotemark{a} & 06/23/95 & 3600 & 20.61 & 19.07 \nl
 KKC 70  & 06/25/95 & 3360 & 20.66 & 18.76 \nl
1331+170 & 05/20/95 & 3600 & 20.85 & 19.24 \nl
1331+2808 & 03/13/96 & 3600 & 20.72 & 19.28 \nl
1344+2818 & 03/12/96 & 2700 & 20.62 & 19.08 \nl
1502+105 & 05/20/95 & 3600 & 20.93 & 19.41 \nl
1540+110 & 05/20/95 & 3600 & 20.50 & 18.91 \nl
1608+4636 & 05/22/95 & 3600 & 20.76 & 19.13 \nl
1634+706 & 05/20/95 & 3600 & 20.73 & 19.09 \nl
1700+4744 & 05/22/95 & 4500 & 20.85 & 18.93 \nl
2145+067 & 08/04/95 & 3600 & 20.56 & 18.41 \nl
2350-0132 & 08/05/95 & 3600 & 20.65 & 18.67 \nl
2354+0048 & 08/05/95 & 3450 & 20.90 & 18.43 \nl
2358+0038 & 08/04/95 & 3600 & 20.95 & 18.89

\enddata
\tablenotetext{a}{Close pair of Q1306+296 and Q1307+296 suggested
surveyed area inbetween.  This field is halfway between the two 
QSOs.} 
\label {tab:  z1obs}
\end{deluxetable}
\renewcommand{\arraystretch}{1.0}

\clearpage
\renewcommand{\arraystretch}{.5}
\begin{deluxetable}{lcccc}

\tablecolumns{5}
 
\tablecaption{Broad Band Photometry of QSOs}
\tablehead{
\colhead{QSO}&
\colhead{z$_{em}$}&
\colhead{optical\tablenotemark{a}}&
\colhead{J}&
\colhead{\kp} 
}

\startdata

0107-025 & 0.958 & 18.11 & 15.99 & 15.33 \nl 
0122-003 & 1.007 & 17.0 & 15.78 & 14.50 \nl
0246+009 & 0.953 & 18.77 & 17.24 & 16.52 \nl
0308+002 & 0.955 & 20.49 & 19.8 & 17.4 \nl 
0333+321 & 1.258 & 17.5 & 15.20 & 14.19 \nl
0447-092 & 0.946 & 18.5 & 16.61 & 15.33 \nl
PKS 0500+019 & 1.000 & 21.2 & 14.81 & 14.50 \nl
4c2221   & 0.951 & 19.5 & 17.49 & 15.73 \nl
0743-006 & 0.994 & 17.1 & 12.10 & 10.45 \nl
CBS 0076  & 0.945 & 17 & 15.54 & 14.83 \nl
PC 0903+474 & 0.948 & 18.61 & 17.15 & 16.31 \nl
1050+542 & 0.995 & 18.2 & 18.00 & 16.84 \nl
LBQS 1231+1627 & 0.999 & 18.80 & 17.09 & 17.30 \nl
LBQS 1235+1123 & 0.947 & 18.10 & 16.53 & 15.60 \nl
1244+324 & 0.949 & 17.2 & 16.34 & 15.40 \nl
1331+170 & 2.084 & 16.71 & 15.19 & 13.77 \nl
CCS 1331+2808 & 0.993 & 19.8 & 17.73 & 16.72 \nl
CCS 1344+2818 & 0.992 & 19.4 & 17.51 & 16.47 \nl
1502+105 & 1.00 & 17.79 & 16.59 & 15.56 \nl
1540+110 & 0.992 & 18 & 17.70 & 16.67 \nl
PC 1608+4636 & 0.951 & 19.15 & 17.93 & 16.57 \nl
1634+706 & 1.334 & 14.90 & 13.66 & 12.56 \nl
PC 1700+4744 & 0.994 & 18.07 & 16.187 & 15.89 \nl
2145+067 & 0.990 & 16.47 & 14.44 & 13.52 \nl
LBQS 2350-0132 & 0.993 & 17.39 & 15.334 & 14.76 \nl
LBQS 2354+0048 & 0.999 & 18.54 & 16.80 & 15.83 \nl
LBQS 2358+0038 & 0.949 & 18.73 & 16.54 & 15.72

\enddata
\tablenotetext{a}{Optical magnitudes were taken from
the NED catalog.  Typically they correspond to B or V} 
\label {tab:  z1 qso phot}
\end{deluxetable}
\renewcommand{\arraystretch}{1.0}

\clearpage
\renewcommand{\arraystretch}{.5}
\begin{deluxetable}{lllll}

\tablecolumns{5}
 
\tablecaption{Narrow Band Observations of $z\sim 1$ Targets}
\tablehead{
\colhead{Field}&
\colhead{filter}&
\colhead{date}&
\colhead{itime (sec)}&
\colhead{1$\sigma$ SFR (\myr)}
}

\startdata
0107-025 & \pb    & 10/20/96 & 14400 & 6 \nl
0122-003 & \pb    & 10/21/96 & 14580 & 6 \nl
0500+019 & 1.3095 & 10/20-21,23/96\tablenotemark{a} & 10620 & 8 \nl
1244+324 & \pb    & 05/22/95 & 11520 &  10 \nl
1331+170 & \pb    & 05/06/96 & 14580 & 6 \nl
1608+4636 & \pb   & 05/05-6/96 & 14580 & 6 \nl
1634+706 & 1.3095 & 08/06/95 & 11340 & 8 \nl
1700+4744 & 1.3095 & 08/07/95 & 12960 & 4 \nl
2145+067 & 1.3095 & 10/20-21/96 & 11340 & 10 \nl
2350-0132 & 1.3095 & 08/09/95 & 13500 & 6 \nl
2354+0048 & 1.3095 & 08/05/95 & 12960 & 4 \nl
2358+0038 & \pb   & 08/08/95 & 13860 & 6

\enddata
\label {tab:  z1nbobs}
\end{deluxetable}
\renewcommand{\arraystretch}{1.0}

\clearpage
\renewcommand{\arraystretch}{.5}
\begin{deluxetable}{lllll}

\tablecolumns{5}
 
\tablecaption{Observations of $z=1.5$ Targets}
\tablehead{
\colhead{Field}&
\colhead{date}&
\colhead{itime H (sec)}&
\colhead{itime n.b.}&
\colhead{H 5$\sigma$} 
}

\startdata

87GB 1554+3526  &  01/30/94     &  600   & 2000 & 21.0  \nl
LBQS 2236-0023  &  07/18-19/94  &  1020  & 3240 & 21.5  \nl
CSO 190\tablenotemark{a}    &  01/28/99  & 1680   &  4320    &  21.9 \nl  
Q 1038+311\tablenotemark{a} &  01/28/99  & 1620   &  2640    &  21.3  \nl  
Q 1147+339      &  01/28/99     &  840   & 2400 & 20.6  \nl
Q 1316+3111     &  01/28/99     &  1080  & 2400 & 21.0 \nl

\enddata
\tablenotetext{a}{Due to observing constraints, these objects were chosen as
signposts, even though they lie more than 1\%~from the center of the narrow band
filter (both at $z=1.45$).}
\label {tab:  z1pt5nbobs}
\end{deluxetable}
\renewcommand{\arraystretch}{1.0}

\clearpage
\renewcommand{\arraystretch}{.5}
\begin{deluxetable}{lllllll}

\tablecolumns{7}
 
\tablecaption{Narrow Band Detections at $z\sim 1$}
\tablehead{
\colhead{Field}&
\colhead{z}&
\colhead{$\Delta$m}&
\colhead{I}&
\colhead{J}&
\colhead{\kp}&
\colhead{SFR} 
}

\startdata
2145+067 & $\sim$ 0.995  & 1.2 & :22.5 \tablenotemark{a}  & 19.2 & 18.6 & 52.8 \nl
2350-0132 & 0.990 & 0.9 & 21.9 & 20.2 & 19.1 & 26.0

\enddata
\tablenotetext{a}{magnitude is uncertain due to overlap with PSF of bright
star}
\label {tab:  z1 nb detections}
\end{deluxetable}
\renewcommand{\arraystretch}{1.0}

\clearpage
\renewcommand{\arraystretch}{.5}
\begin{deluxetable}{lllllll}

\tablecolumns{7}
 
\tablecaption{Comparison of Narrow-Band Surveys}
\tablehead{
\colhead{redshift}&
\colhead{density}&
\colhead{$3\sigma$~flux limit}&
\colhead{$<SFR>$}&
\colhead{Number}&
\colhead{reference}&
\colhead{notes} \\
\colhead{} &
\colhead{$10^{-4}$~Mpc$^{-3}$} &
\colhead{$10^{-16}$}&
\colhead{M$_{\odot}$/yr}&
\colhead{of} &
\colhead{}&
\colhead{on field} \\
\colhead{}&
\colhead{(comoving)}&
\colhead{ergs/cm$^2$/s}&
\colhead{}&
\colhead{Galaxies} &
\colhead{}&
\colhead{selection}
}

\startdata

0.7--1.9       &   $2 $   &  0.6  &  50                  & 30 & McCarthy 1999            & random \tablenotemark{a} \nl
0.95--1.0      &   $4.7$  &  1.3  &  35                  & 2  & this work                & QSO em. \nl
1.5            &   $<40$  &  :5.0 &  \nodata             & 0  & this work                & QSO em. \nl
2.0--2.7       &   $<70$  &  10   &  \nodata             & 0  & Bunker 1999              & DLA \tablenotemark{d}\nl
2.2--2.3       &   $22.5$ &  1.9  &  $\sim 40$           & 2  & van der Werf 1997        & radio galaxies \nl
2.2--2.5       &   $0.33$ &  3.4  &  250\tablenotemark{b}& 1  & Thompson 1996            & QSO em. \nl
2.28--2.29     &   $<120$ &  0.9  &  \nodata             & 0  & Pahre 1995               & QSO em. \nl
2.3--2.4,0.89  &   $9$    &  4.8  &  70                  & 18 & MTBW98                   & Abs. line \nl
2.3--2.5       &   60     &  1.0  &  50                  & 5  &TMM98\tablenotemark{c}    & Abs. line  \nl

\enddata
\tablenotetext{a}{Slitless spectroscopy}
\tablenotetext{b}{See Beckwith et al. 1998 for a discussion}
\tablenotetext{c}{We have counted the density of objects from that survey with line fluxes greater
than $1\times 10^{-16}$~ergs/cm$^2$/s.}
\tablenotetext{d}{longslit spectroscopy}
\label {tab:  halpha_surveys_tab}
\end{deluxetable}
\renewcommand{\arraystretch}{1.0}

\clearpage

\renewcommand{\arraystretch}{.5}
\begin{deluxetable}{llll}
\tablecolumns{4}

\tablecaption{J-band Number-Magnitude Counts}

\tablehead{
\colhead{J mag.}&
\colhead{N/deg$^2$/mag.}&
\colhead{lower limit $1 \sigma$}&
\colhead{upper limit $1 \sigma$}
}

\startdata

      15.7500   &   155.153  &     80.9122   &    277.840  \nl
      16.2500   &   310.305  &     202.940   &    463.519  \nl
      16.7500   &   659.399  &     501.143   &    861.098  \nl
      17.2500   &   775.764  &     603.932   &    991.038  \nl
      17.7500   &   1318.80  &     1093.83   &    1586.05  \nl
      18.2500   &   2404.87  &     2099.45   &    2710.29  \nl
      18.7500   &   3646.09  &     3270.02   &    4022.15  \nl
      19.2500   &   6206.11  &     5715.47   &    6696.75  \nl
      19.7500   &   9301.09  &     8689.90   &    9912.28  \nl
      20.2500   &   13702.9  &     12907.2   &    14498.5  \nl
      20.7500   &   17970.0  &     17013.4   &    18926.6  \nl
      21.2500   &   27714.0  &     26278.3   &    29149.8  \nl
      21.7500   &   69312.5  &     64653.9   &    73971.2  \nl

\enddata
\label{tab: jnc}
\end{deluxetable}
\renewcommand{\arraystretch}{1.0}

%%% Local Variables: 
%%% mode: plain-tex
%%% TeX-master: t
%%% TeX-master: t
%%% End: 

\clearpage 

%\figcaption[]{}
\figcaption[]{J-band image of the 0107-025 field. \label{fig:  im0107}}

\figcaption[]{J-band number counts for the 0107-025 field.  The error
  bars show the number counts for this field.  The squares are the
  average of all J-band observations of z=1 fields \label{fig:  comphist0107}}

\figcaption[]{J-band image of the 2145+067 field.  The \ha-emitter is
circled and label ``A''.  \label{fig:  im2145}}

\figcaption[]{J-band number counts for the 2145+067 field.  The square
  points are the average of all J-band observations of z=1 fields
  \label{fig: comphist2145}}

\figcaption[]{Broad-Narrow band color magnitude diagram for the
  2145+067 field.  The curved lines indicate the 99.5\% confidence
  interval.  \label{fig:  cm2145}}

\figcaption[]{J-band image of the 2350-032 field. The \ha-emitter is
circled and label ``A''.  \label{fig:  im2350}} 

\figcaption[]{Broad-Narrow band color magnitude diagram for the
  2350-032 field.  The curved lines indicate the 99.5\% confidence
  interval.  \label{fig:  cm2350}}

\figcaption[]{J-band number counts for the 2350-032 field.  The square
  points are the average of all J-band observations of z=1 fields.
  \label{fig: comphist2350}}

\figcaption[]{LRIS spectrum of the candidate object in 2350-0132.
  \label{fig: spec2350}}

\figcaption[]{J-band number-magnitude counts for survey fields.  
  The square points with error bars are the observed number counts.
  The + symbols are the counts from Bershady et al. 1998, and the
  solid line is the fit to the $J<15$~counts presented in Mamon et al.
  1998.  \label{fig:  jnc}}

\figcaption[]{K-band number-magnitude counts for survey fields.  
  The square points with error bars are the observed number counts.
  The + symbols are the counts from the literature as compiled in
  Gardner 1998. \label{fig:  knc} }

\figcaption[]{ Number of galaxies detected in the J band in the range
  $18<J<22$~with at least 3$\sigma$~confidence in radial bins
  5\arcs~wide.  The error bars are the Poissonian error on the number
  of galaxies detected in each bin.  The dotted line is the average
  number of galaxies per square arcminute in the total number counts
  in the same magnitude range. \label{fig: radhist} }

\clearpage

\addtocounter{figure}{-12}

\begin{figure}
%\plotone{im0107.eps}
\plotone{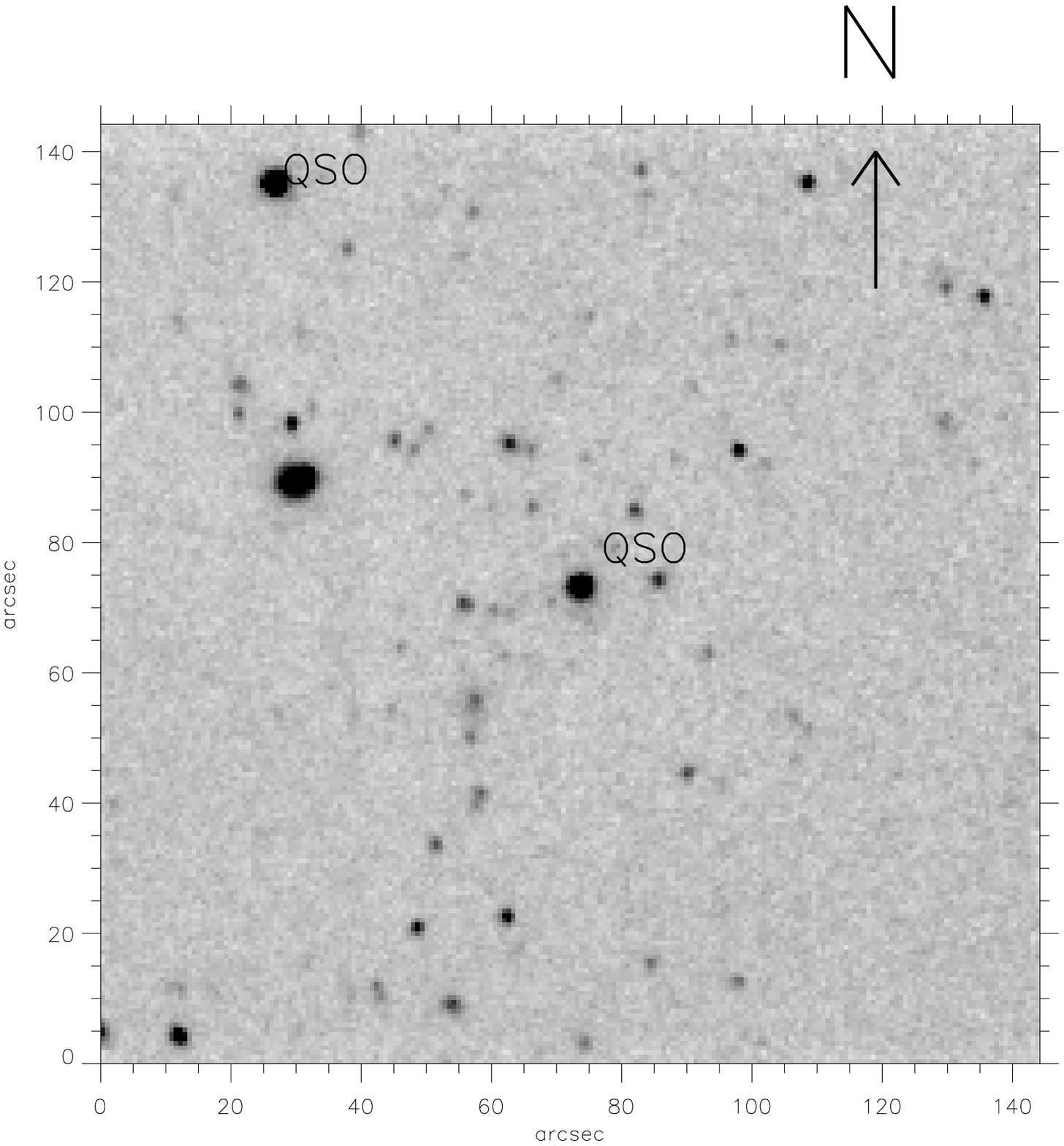}
%\caption{J-band image of the 0107-025 field.}
%\label{fig:  im0107}
\end{figure}

\begin{figure}
%\plotone{jnc0107.eps}
\plotone{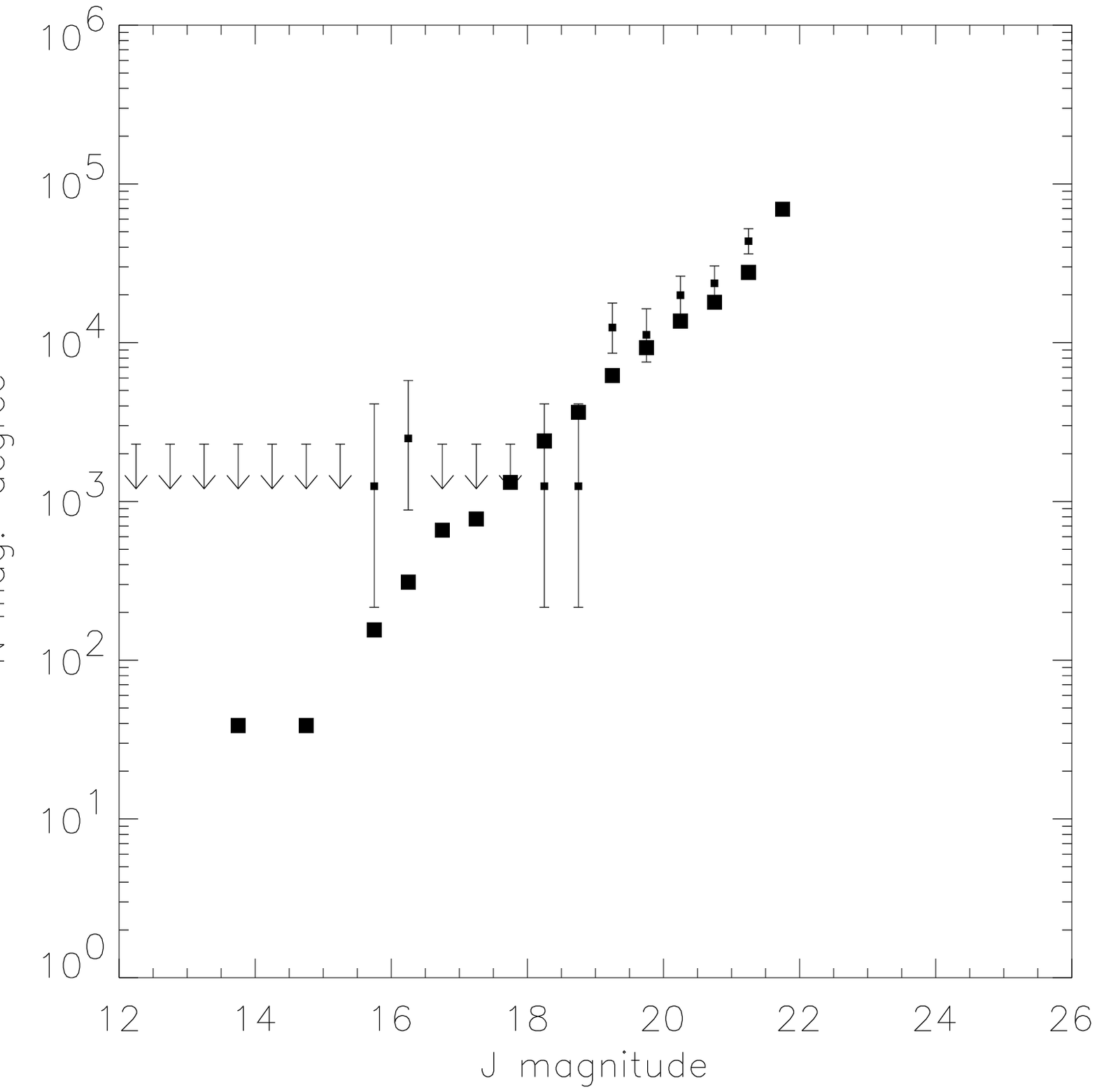}
%\caption{J-band number counts for the 0107-025 field.  The error bars show 
%  the number counts for this field.  The sqaures are the average of
%  all J-band observations of z=1 fields}
%\label{fig:  comphist0107}
\end{figure}

\clearpage
\begin{figure}
%\plotone{im2145.eps}
\plotone{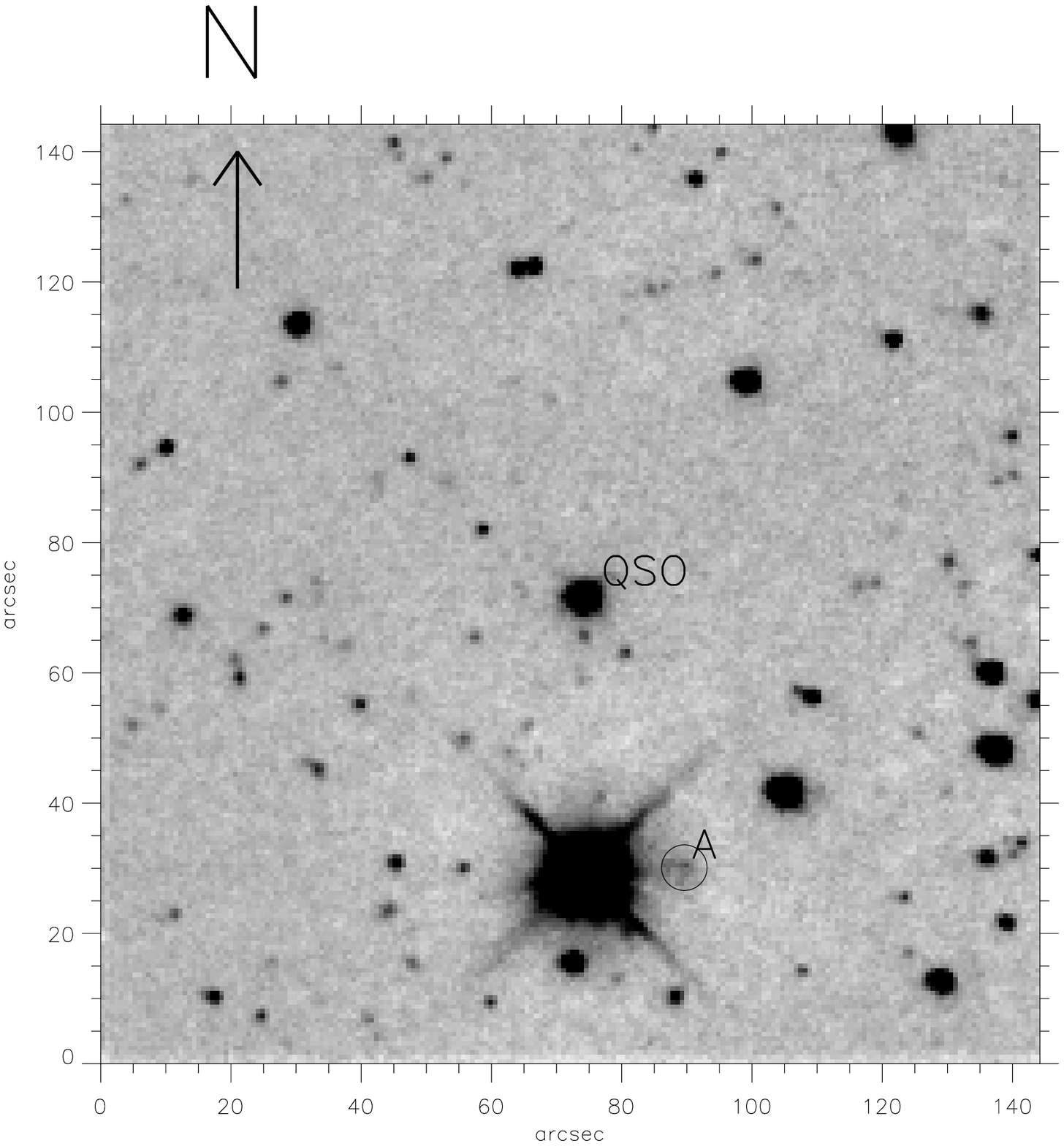}
%\caption{J-band image of the 2145+067 field.  The \ha-emitter is
%circled and label ``A''.}
%\label{fig:  im2145}
\end{figure}

\clearpage
\begin{figure}
%\plotone{jnc2145.eps}
\plotone{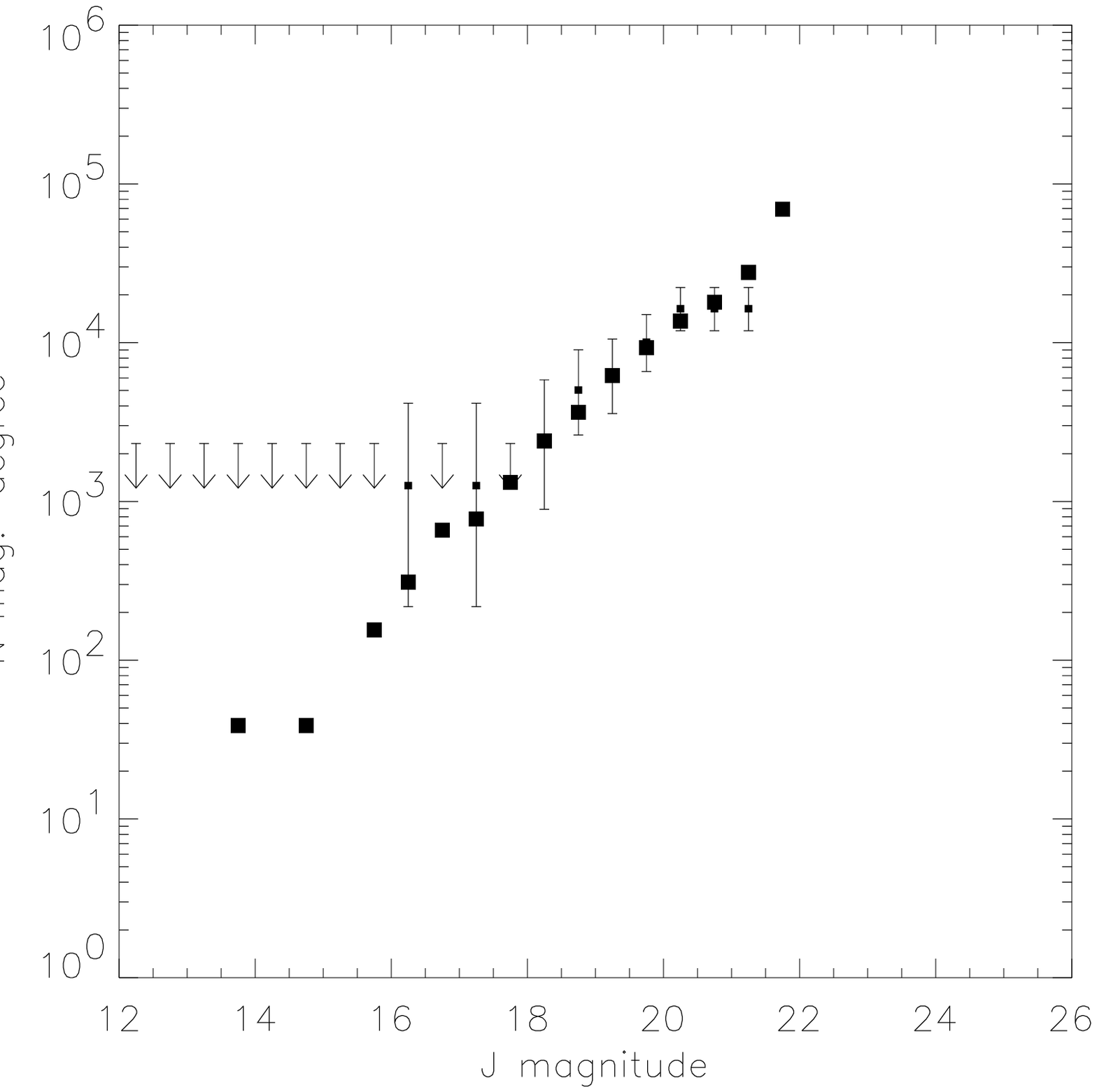}
%\caption{J-band number counts for the 2145+067 field.  The square points
%  are the average of all J-band observations of z=1 fields}
%\label{fig:  comphist2145}
\end{figure}

\clearpage
\begin{figure}
%\plotone{cm2145.eps}
\plotone{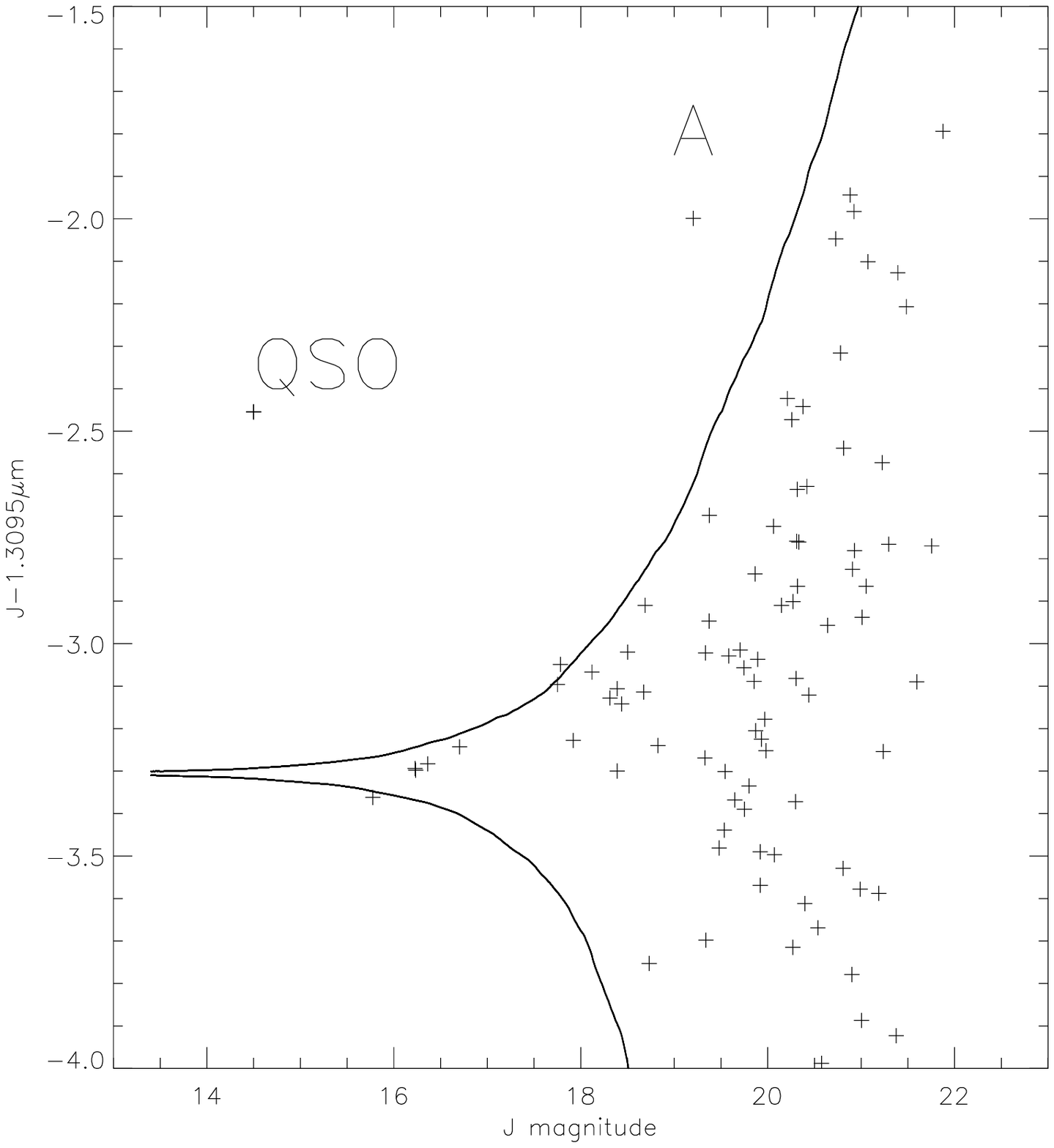}
%\caption{Broad-Narrow band color magnitude diagram for the 2145+067 field. 
%  The curved lines indicate the 99.5\% confidence interval.}
%\label{fig:  cm2145}
\end{figure}

\clearpage
\begin{figure}
%\plotone{im2350.eps}
\plotone{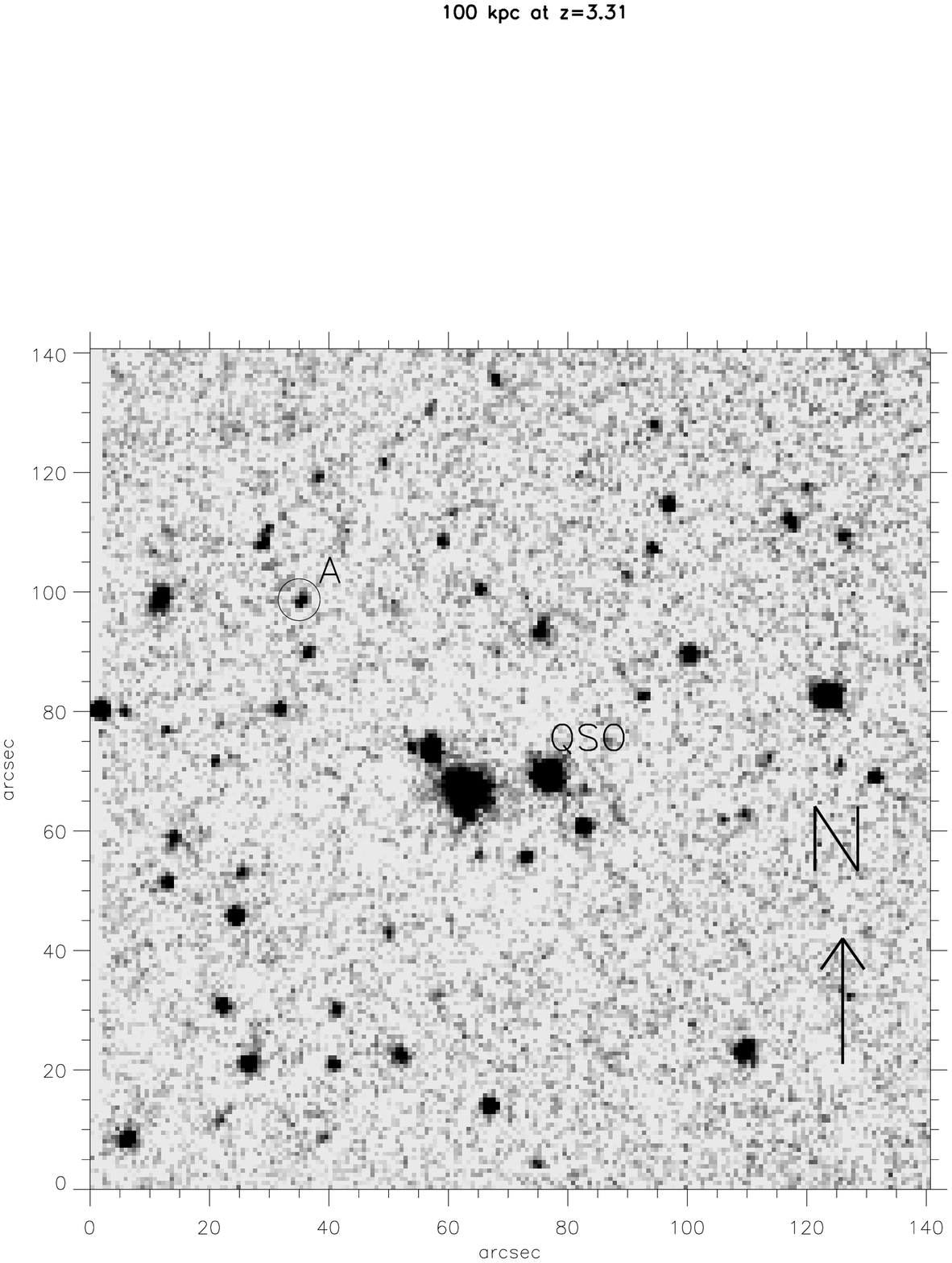}
%\caption{J-band image of the 2350-032 field. The \ha-emitter is
%circled and label ``A''.} 
%\label{fig:  im2350}
\end{figure}

\clearpage

\begin{figure}
%\plotone{cm2350.eps}
\plotone{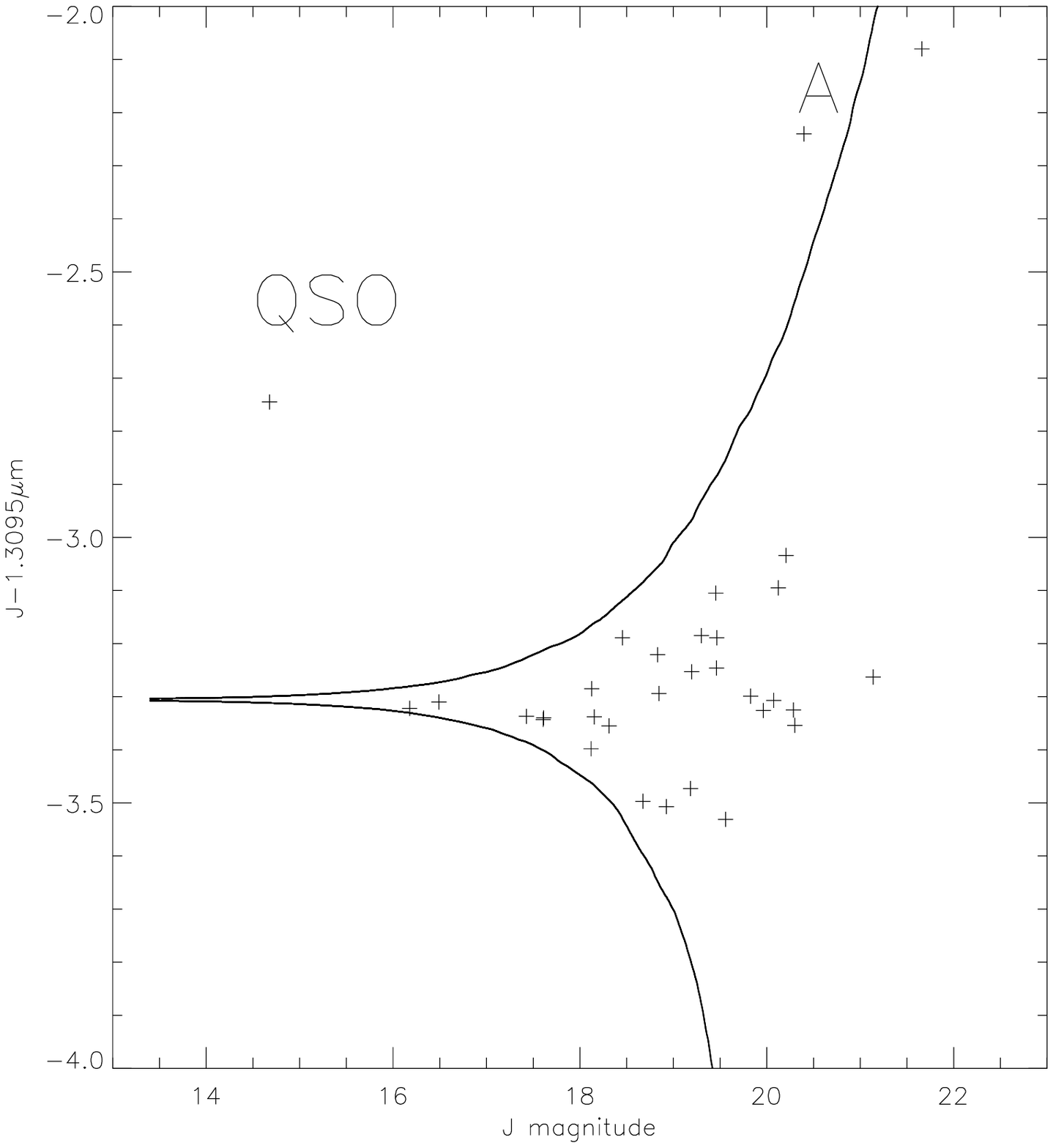}
%\caption{Broad-Narrow band color magnitude diagram for the 2350-032 field. 
%  The curved lines indicate the 99.5\% confidence interval.}
%\label{fig:  cm2350}
\end{figure}

\clearpage

\begin{figure}
%\plotone{jnc2350.eps}
\plotone{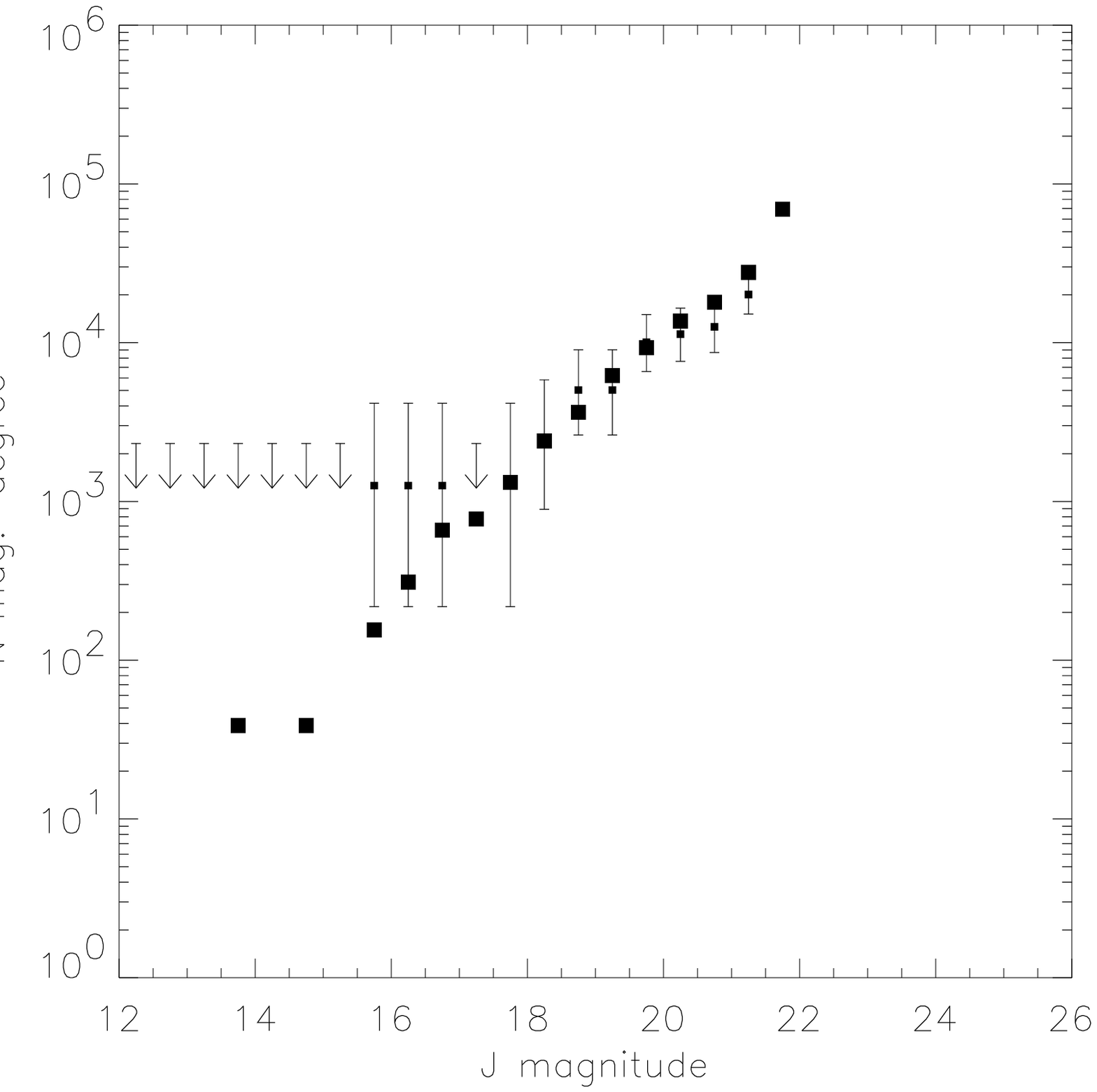}
%\caption{J-band number counts for the 2350-032 field.  The square points 
%  are the average of all J-band observations of z=1 fields}
%\label{fig:  comphist2350}
\end{figure}

\clearpage

\begin{figure}
%\plotone{spec2350.eps}
\plotone{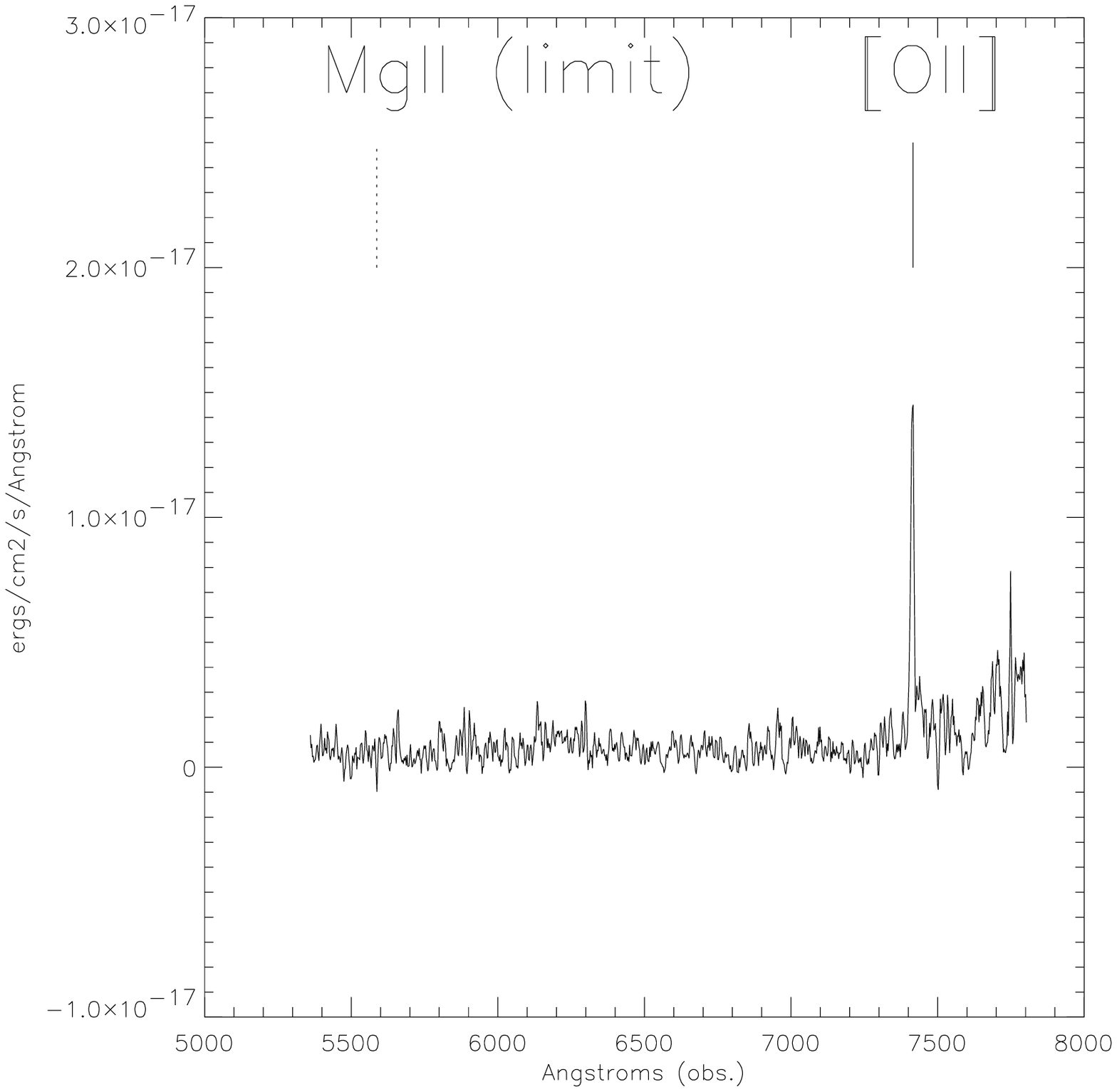}
%\caption{LRIS spectrum of the candidate object in 2350-0132}
%\label{fig:  spec2350}
\end{figure}

\clearpage

\begin{figure}
%\plotone{jnc.eps}
\plotone{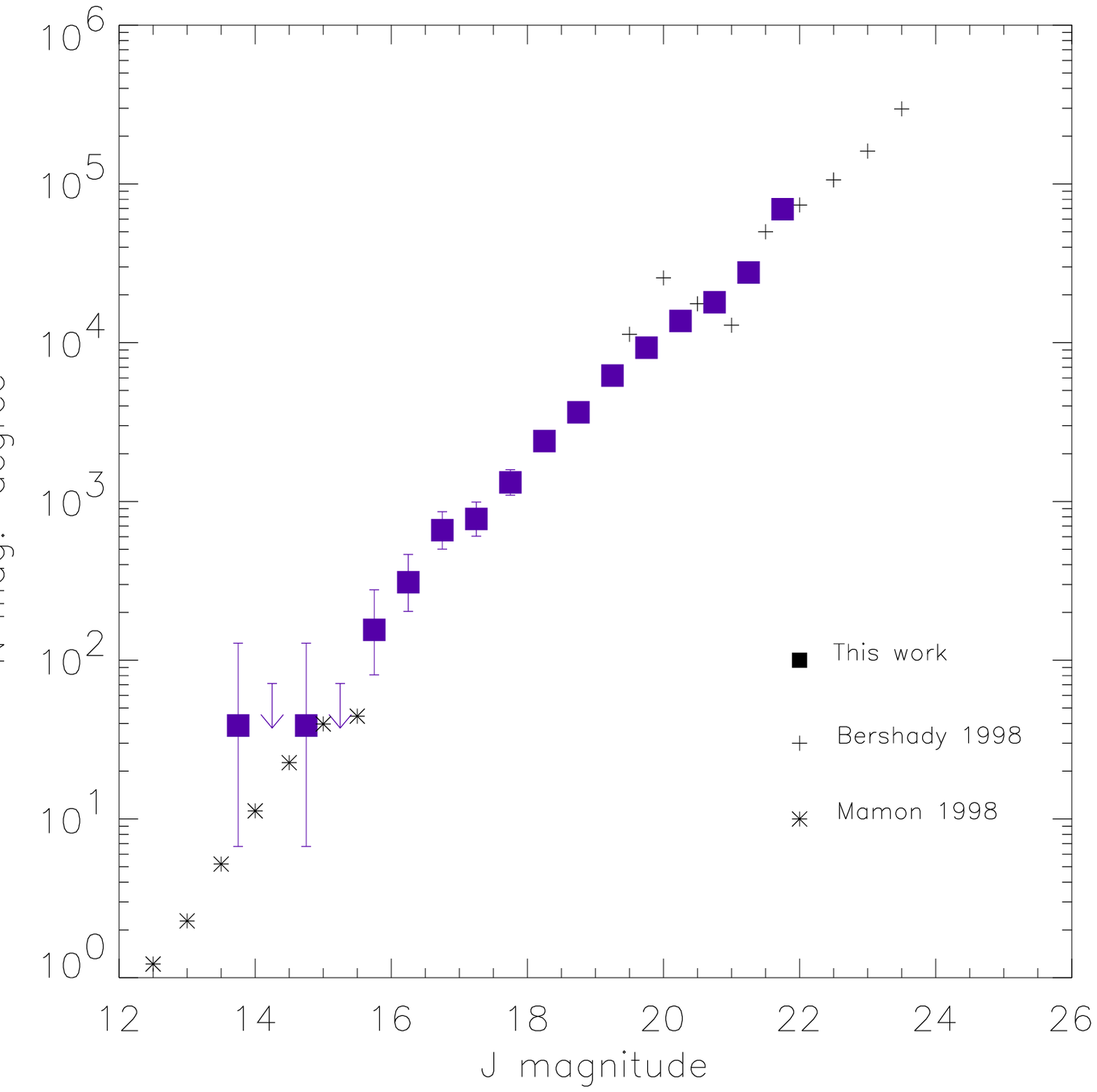}
%\caption{J-band number-magnitude counts for survey fields.  
%  The square points with error bars are the observed number counts.
%  The + symbols are the counts from Bershady et al. 1998, and the
%  solid line is the fit to the $J<15$~counts presented in Mamon et al.
%  1998}
%\label{fig:  jnc}
\end{figure}

\clearpage

\begin{figure}
%\plotone{knc.eps}
\plotone{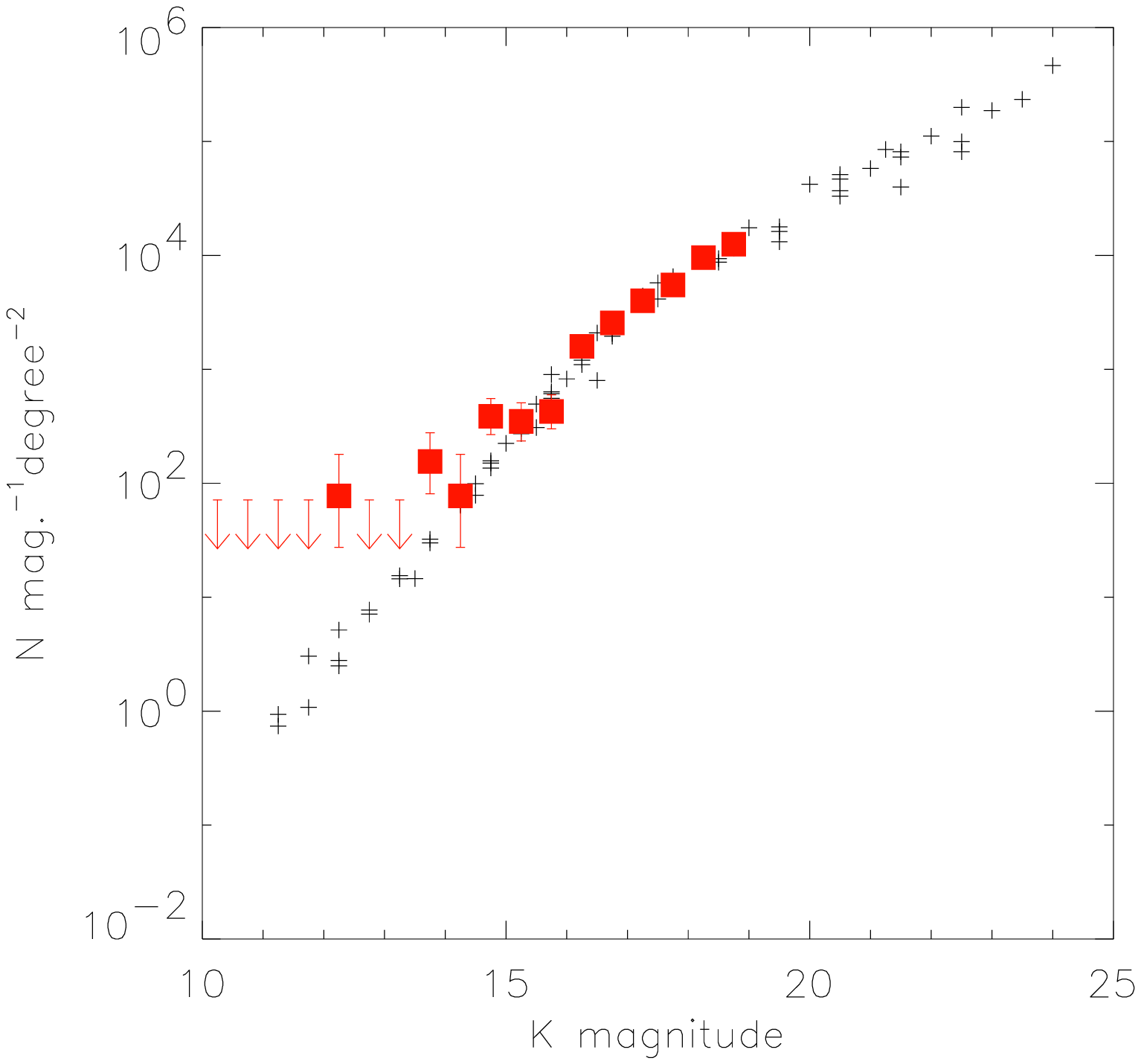}
%\caption{K-band number-magnitude counts for survey fields.  
%  The square points with error bars are the observed number counts.
%  The + symbols are the counts from the literature as compiled in
%  Gardner 1998.}
%\label{fig:  knc}
\end{figure}

\begin{figure}
%\plotone{radhist.eps}
\plotone{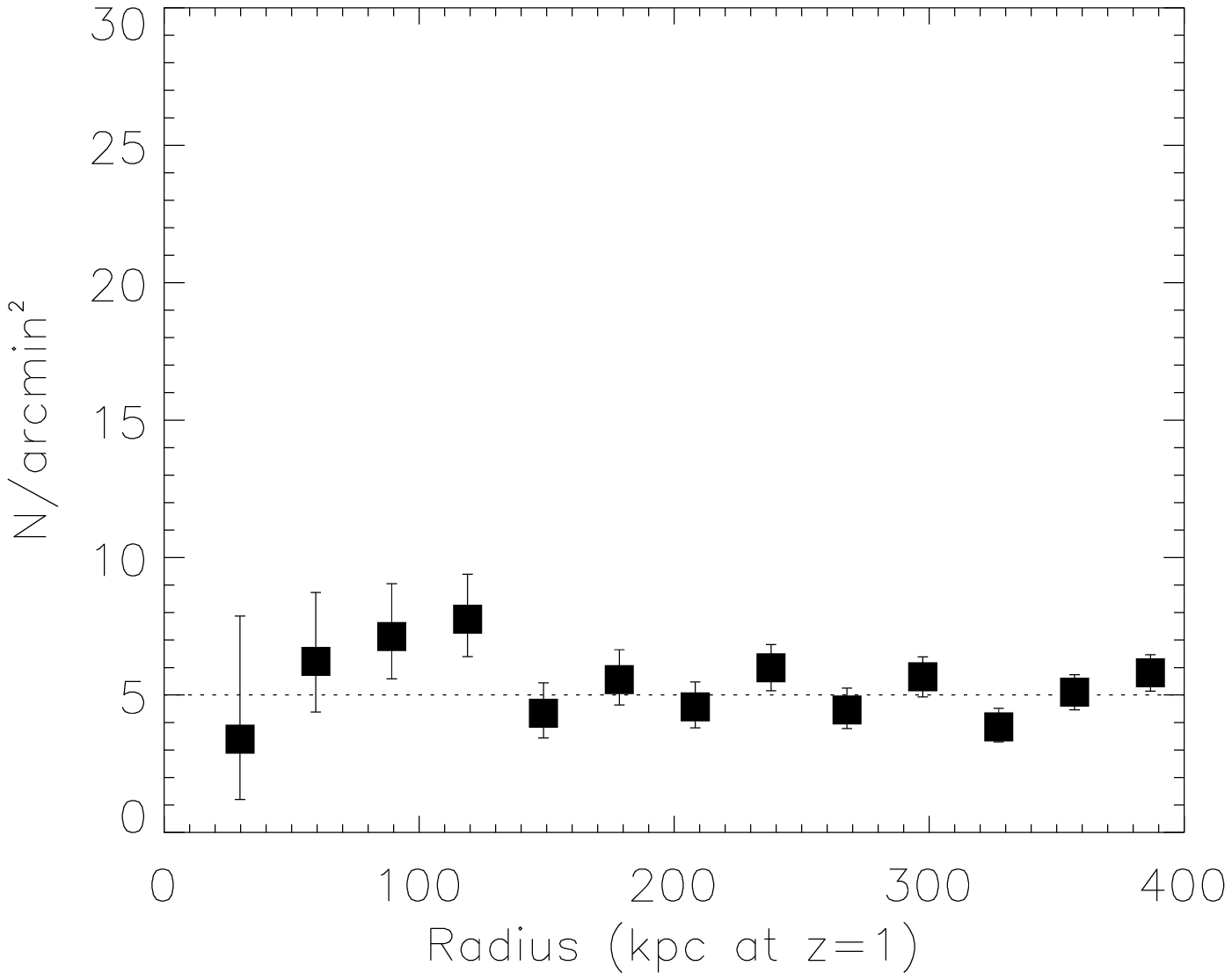}
%\caption{
%  Number of galaxies detected in the J band in the range
%  $18<J<22$~with at least 3$\sigma$~confidence in radial bins
%  5\arcs~wide.  The error bars are the Poissonian error on the number
%  of galaxies detected in each bin.  The dotted line is the average
%  number of galaxies per square arcminute in the total number counts
%  in the same magnitude range.
%}
%\label{fig:  radhist}
\end{figure}

\end{document}